\documentclass{mn2e}
\usepackage{epsfig}
\newread\epsffilein    
\newif\ifepsffileok    
\newif\ifepsfbbfound   
\newif\ifepsfverbose   
\newdimen\epsfxsize    
\newdimen\epsfysize    
\newdimen\epsftsize    
\newdimen\epsfrsize    
\newdimen\epsftmp      
\newdimen\pspoints     
\def\epsfbox#1{\global\def\epsfllx{72}\global\def\epsflly{72}%
   \global\def\epsfurx{540}\global\def\epsfury{720}%
   \def\lbracket{[}\def\testit{#1}\ifx\testit\lbracket
   \let\next=\epsfgetlitbb\else\let\next=\epsfnormal\fi\next{#1}}%
\def\epsfgetlitbb#1#2 #3 #4 #5]#6{\epsfgrab #2 #3 #4 #5 .\\%
   \epsfsetgraph{#6}}%
\def\epsfnormal#1{\epsfgetbb{#1}\epsfsetgraph{#1}}%
\def\epsfgetbb#1{%
\openin\epsffilein=#1
\ifeof\epsffilein\errmessage{I couldn't open #1, will ignore it}\else
   {\epsffileoktrue \chardef\other=12
    \def\do##1{\catcode`##1=\other}\dospecials \catcode`\ =10
    \loop
       \read\epsffilein to \epsffileline
       \ifeof\epsffilein\epsffileokfalse\else
          \expandafter\epsfaux\epsffileline:. \\%
       \fi
   \ifepsffileok\repeat
   \ifepsfbbfound\else
    \ifepsfverbose\message{No bounding box comment in #1; using defaults}\fi\fi
   }\closein\epsffilein\fi}%
\def\epsfsetgraph#1{%
   \epsfrsize=\epsfury\pspoints
   \advance\epsfrsize by-\epsflly\pspoints
   \epsftsize=\epsfurx\pspoints
   \advance\epsftsize by-\epsfllx\pspoints
   \epsfxsize\epsfsize\epsftsize\epsfrsize
   \ifnum\epsfxsize=0 \ifnum\epsfysize=0
      \epsfxsize=\epsftsize \epsfysize=\epsfrsize
     \else\epsftmp=\epsftsize \divide\epsftmp\epsfrsize
       \epsfxsize=\epsfysize \multiply\epsfxsize\epsftmp
       \multiply\epsftmp\epsfrsize \advance\epsftsize-\epsftmp
       \epsftmp=\epsfysize
       \loop \advance\epsftsize\epsftsize \divide\epsftmp 2
       \ifnum\epsftmp>0
          \ifnum\epsftsize<\epsfrsize\else
             \advance\epsftsize-\epsfrsize \advance\epsfxsize\epsftmp \fi
       \repeat
     \fi
   \else\epsftmp=\epsfrsize \divide\epsftmp\epsftsize
     \epsfysize=\epsfxsize \multiply\epsfysize\epsftmp   
     \multiply\epsftmp\epsftsize \advance\epsfrsize-\epsftmp
     \epsftmp=\epsfxsize
     \loop \advance\epsfrsize\epsfrsize \divide\epsftmp 2
     \ifnum\epsftmp>0
        \ifnum\epsfrsize<\epsftsize\else
           \advance\epsfrsize-\epsftsize \advance\epsfysize\epsftmp \fi
     \repeat     
   \fi
   \ifepsfverbose\message{#1: width=\the\epsfxsize, height=\the\epsfysize}\fi
   \epsftmp=10\epsfxsize \divide\epsftmp\pspoints
   \newcount\figskipcount
      \message{#1 \the\epsfysize  }
   \vbox to\epsfysize{\vfil\hbox to\epsfxsize{%
      \includegraphics{#1}%
      \hfil}}%
\epsfxsize=0pt\epsfysize=0pt}%

{\catcode`\%=12 \global\let\epsfpercent=
\long\def\epsfaux#1#2:#3\\{\ifx#1\epsfpercent
   \def\testit{#2}\ifx\testit\epsfbblit
      \epsfgrab #3 . . . \\%
      \epsffileokfalse
      \global\epsfbbfoundtrue
   \fi\else\ifx#1\par\else\epsffileokfalse\fi\fi}%
\def\epsfgrab #1 #2 #3 #4 #5\\{%
   \global\def\epsfllx{#1}\ifx\epsfllx\empty
      \epsfgrab #2 #3 #4 #5 .\\\else
   \global\def\epsflly{#2}%
   \global\def\epsfurx{#3}\global\def\epsfury{#4}\fi}%
\def\epsfsize#1#2{\epsfxsize}
\def\figinsert#1#2{\epsfbox{#1} \message{#2} }
\def\spm{\mbox{\scriptsize{ $\pm$ }}}
\voffset-.8in

\title[Mid-infrared sources in the ELAIS Deep X-ray Survey]
  {Mid-infrared sources in the ELAIS Deep X-ray Survey}
\author[J.C. Manners, et al.]
  {J.C. Manners$^1$, S. Serjeant$^{2,3}$, S. Bottinelli$^4$, M. Vaccari$^{1,2,5}$, A. Franceschini$^{1}$, \and I. Perez-Fournon$^{6}$, E. Gonzalez-Solares$^{7}$, C.J. Willott$^8$, O. Johnson$^9$, O. Almaini$^{10}$, \and M. Rowan-Robinson$^{2}$, S. Oliver$^{11}$\\
\\
 $^1$Dipartimento di Astronomia, Universit\`a di Padova, Vicolo dell'Osservatorio 2, I-35122, Padova, Italy\\ 
 $^2$Astrophysics Group, Blackett Laboratory, Imperial College, Prince Consort Road, London SW7 2BW, UK\\ 
 $^3$Centre for Astrophysics and Planetary Science, School of Physical Sciences, University of Kent, Canterbury, Kent, CT2 7NR, UK\\
 $^4$Institute for Astronomy, University of Hawaii, 2680 Woodlawn Drive, Honolulu, HI 96822, USA\\
 $^5$CISAS "G. Colombo", Universit\`a di Padova, Via Venezia 15, I-35131, Padova, Italy\\
 $^6$Instituto de Astrofisica de Canarias, C/ Via Lactea, 38200 La Laguna, S/C de Tenerife, Spain\\
 $^7$Institute of Astronomy, Madingley Road, Cambridge, CB3 0HA, UK\\
 $^8$Herzberg Institute of Astrophysics, National Research Council, 5071 West Saanich Rd, Victoria, B.C. V9E 2E7, Canada\\
 $^9$Institute for Astronomy, University of Edinburgh, Royal Observatory, Blackford Hill, Edinburgh EH9 3HJ, UK\\
 $^{10}$School of Physics and Astronomy, University of Nottingham, University Park, Nottingham NG7 2RD, UK\\
 $^{11}$Astronomy Centre, Department of Physics \& Astronomy, University of Sussex, Brighton, BN1 9QJ, UK\\
}
\date{MNRAS in press}
\pubyear{2004}

\begin{document}

\label{firstpage}

\maketitle

\begin{abstract}
We present a cross-correlation of the European Large Area {\it ISO} survey
(ELAIS) with the ELAIS Deep X-ray Survey of the N1 and N2 fields.
There are 7 {\it Chandra} point sources with matches in the ELAIS Final
Analysis  15$\mu$m catalogue, out of a total of 28 extragalactic {\it ISO}
sources present in the {\it Chandra} fields. Five of these are consistent
with AGN giving an AGN fraction of $\sim 19$ per cent in the
$15\mu$m flux range $0.8 - 6$ mJy. We have co-added
the hard X-ray fluxes of the individually-undetected {\it ISO} sources and
find a low significance detection consistent with star formation in
the remaining population. We 
combine our point source cross-correlation fraction with the {\it XMM-Newton}
observations of the Lockman Hole and {\it Chandra} observations of the 
Hubble Deep Field North to constrain source count models 
of the mid-infrared galaxy population. 
The low dust-enshrouded AGN fraction in ELAIS implied by
the number of cross-identifications between the ELAIS 
mid-infrared sample and the {\it Chandra} point sources is encouraging for
the use of mid-infrared surveys to constrain the cosmic star formation 
history, provided there are not further large undetected populations
of Compton-thick AGN. 
\end{abstract}


\section{Introduction}

Enormous progress has recently been made in resolving the sources that
comprise the extragalactic hard X-ray background (e.g. Mushotzky {\it
et al.} 2000, Cowie {\it et al.} 2002, Moretti {\it et al.}
2003). Studies are now focusing on characterising the nature of these
sources through multi-waveband imaging and spectroscopy. Surveys with
the {\it ISO} satellite have recently discovered strong evolution in
the  mid-infrared galaxy population (e.g. Elbaz {\it et al.} 1999,
Serjeant {\it et al.} 2000, Chary \& Elbaz   2001, Gruppioni {\it et
al.} 2002) which is thought to be mainly due to a strongly evolving
obscured cosmic star formation history (e.g. Aussel {\it et al.}
1999), although there is also a contribution from  dust-enshrouded
AGN. Optical spectroscopic follow-ups of these samples are underway
(e.g. La Franca {\it et al.}  2003, Gonzalez-Solares {\it et al.}
2004, Perez-Fournon {\it et al.} in preparation) but as the sources
might not be optically-thin at optical wavelengths, nor emitting
isotropically, it is possible that heavily dust-enshrouded populations
may be mis-classified by this approach.  The {\it Chandra}
observations of the HDF-North (e.g. Hornschmeier {\it et al.} 2001,
Alexander {\it et al.} 2002) placed constraints on the fraction of
dust-enshrouded AGN at  the faintest end of the $15\mu$m source
counts. Most of the known $15\mu$m galaxies are at somewhat higher
flux densities, however. Some  inroads were made on the AGN fraction
for brighter mid-infrared sources  by Alexander et al. (2001) using
{\it BeppoSAX} observations of the ELAIS survey (described below), but
the  {\it BeppoSAX} depth was not sufficient to detect most
Compton-thin Seyfert II galaxies in the targeted ELAIS field. Fadda et
al. (2002) provided the first study with a reasonably significant
number of sources with hard X-ray and mid-infrared emission. They
combined {\it XMM-Newton} observations of the Lockman Hole and {\it
Chandra} observations of the HDF-N with coincident ISOCAM data. This
paper will provide further statistics using 2 {\it Chandra} pointings
in the ELAIS northern survey regions N1 and N2. An upcoming analysis
of the {\it Spitzer} Wide-area Infrared Extragalactic Survey (SWIRE,
Lonsdale {\it et al.} 2003) observations in the N1 region will provide
enhanced statistics in complementary wavebands.

The European Large Area {\it ISO} Survey, ELAIS, was the largest open time
project on {\it ISO}, covering wavelengths from $6.7\mu$m to $175\mu$m
(Oliver {\it et al.} 2000, Rowan-Robinson {\it et al.} 2004). The survey, 
and its follow-ups, have many
ambitious aims,  including tracing the cosmic star formation history
to $z\sim1$ and the discovery of ultra- and hyper-luminous galaxies at
high redshift.  The mid-infrared source counts (Serjeant {\it et al.} 2000, 
Gruppioni {\it et al.} 2002, Elbaz {\it et al.} in preparation)
and far-infrared  source counts (Efstathiou {\it et al.} 2000) both show
evidence for strong evolution, as does the far-infrared luminosity
function (Serjeant {\it et al.} 2001).  The ELAIS mid-infrared source counts
cover the transition from Euclidean slope to steep evolution; by
virtue of the large area and large investment in observing time, ELAIS
represents by far the largest sample of galaxies from this strongly
evolving mid-infrared population. A high proportion have been revealed as 
ultra-luminous infrared galaxies (14\% of $15\mu$m galaxies with known z)
including 9 hyper-luminous infrared galaxies (Morel {\it et al.} 2001, 
Rowan-Robinson {\it et al.} 2004).

Thanks to extensive multi-wavelength coverage, the ELAIS fields have
now arguably become the most well studied regions of their size, and
natural targets for on-going or planned large-area surveys with the
most powerful ground and space-based facilities. Further details on
ELAIS multi-wavelength observations and catalogues are presented in
Rowan-Robinson {\it et al.} (2004). In particular, ELAIS 15
\mbox{$\mu$m} observations will complement the SWIRE survey in three
areas (N1, N2 and S1) by covering the 8-24um gap in {\it Spitzer}'s filters.

In the ELAIS Deep X-ray survey we made deep {\it Chandra} pointings of
the ELAIS N1 and N2 fields to a limiting depth of $\sim10^{-15}$ erg
s$^{-1}$ cm$^{-2}$ in the $0.5-8$keV band, with a total exposure of
$\sim75$ks in each field. Manners {\it et al.} (2003) present  the
data and source counts. Gonzalez-Solares {\it et al.} (in preparation)
present follow-up imaging  and optical spectroscopy of the sample and
Willott {\it et al.} (2003) present Subaru infrared
spectroscopy. Almaini {\it et al.} (2003) present a  cross-correlation
of the {\it Chandra} sources with the sub-mm sources of Scott {\it et
al.} (2002) and Fox {\it et al.} (2002),  and measure the clustering
of the {\it Chandra} population.

Here we present a cross-correlation of the {\it Chandra} X-ray sources
with the mid-infrared sources from the ELAIS survey. Section
\ref{sec:observations} summarises the data acquisition in the
mid-infrared and X-ray, and provides references to more
exhaustive descriptions for the interested reader. Section
\ref{sec:method} describes the cross-correlation between the {\it ISO} and
{\it Chandra} data, and section \ref{sec:results} discusses the significance
of the results.

\section{Observations}\label{sec:observations}

\begin{figure}
\centering
\centerline{\epsfxsize=8.5 truecm \figinsert{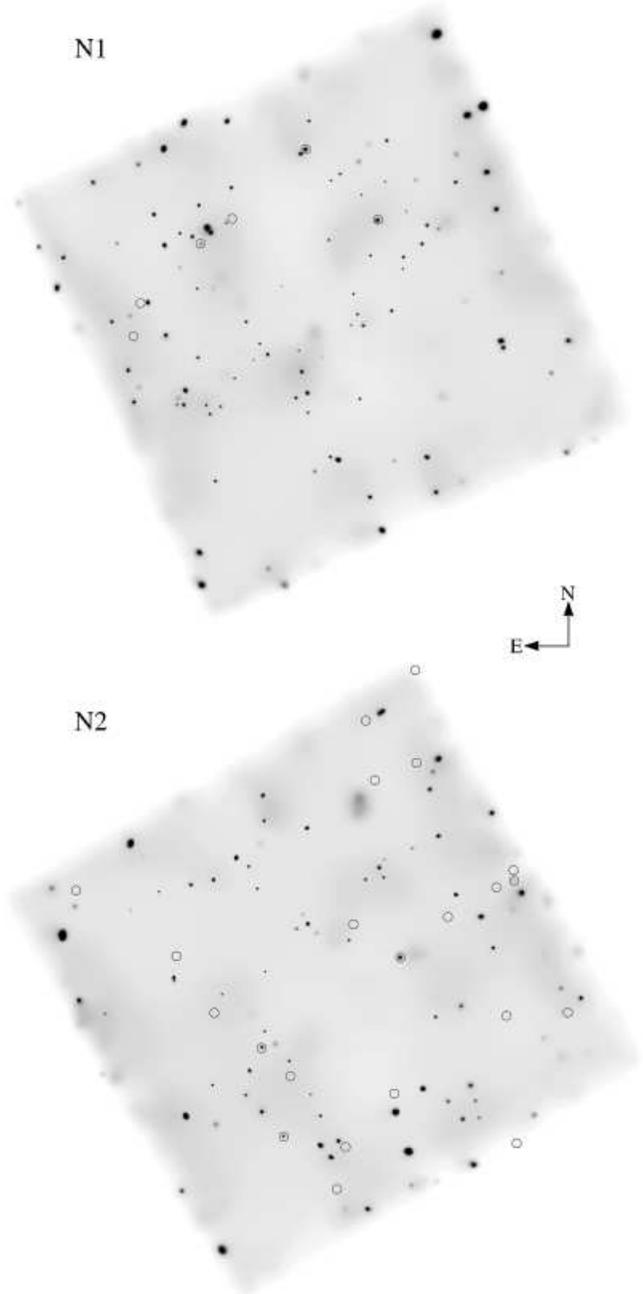}{0.0pt}}
\caption{Positions of the 28 extragalactic 15 $\mu$m {\it ISO} sources
(black circles) superimposed on the smoothed 0.5 - 8 keV {\it Chandra}
images. The images have been adaptively smoothed using the
flux-conserving algorithm CSMOOTH from the Chandra Interactive
Analysis of Observations (CIAO) package.
\label{fig:sources}}
\end{figure}

\begin{table*}
\footnotesize
\begin{tabular}{ccccccccc} \hline
ID$^1$  & {\it Chandra} position$^2$ & {\it ISO} position$^3$ & Offset$^4$ & Prob.$^5$        & $f_{X}$ 0.5 - 8 keV$^6$ & $f_{15\mu m}$$^7$ & z$^8$          & Class$^9$       \\
        & (J2000.0)                  & (J2000.0)              & $''$       & $\times 10^{-3}$ & erg cm$^{-2}$ s$^{-1}$  & mJy               &                &                 \\
\hline
N1\_20  & 16:10:46.57+54:35:38.8     & 16:10:46.66+54:35:39.0 & 2.26       & 1.26             & 2.31 $\times 10^{-15}$  & 1.844             & 0.0634         & SB              \\
N1\_49  & 16:10:20.88+54:39:00.9     & 16:10:20.80+54:39:01.7 & 2.14       & 0.32             & 11.1 $\times 10^{-15}$  & 2.918             &                & AGN 2           \\
N1\_69  & 16:10:03.18+54:36:28.4     & 16:10:03.01+54:36:30.1 & 4.75       & 0.13             & 59.5 $\times 10^{-15}$  & 1.453             & 0.2675         & AGN 1           \\
N2\_25  & 16:36:55.79+40:59:10.5     & 16:36:55.83+40:59:09.2 & 1.46       & 0.12             & 10.2 $\times 10^{-15}$  & 1.017             & 2.61           & AGN 1$^\ddagger$\\
N2\_33  & 16:36:51.69+40:56:00.4     & 16:36:51.61+40:55:59.0 & 2.12       & 0.79             & 3.05 $\times 10^{-15}$  & 1.732             & 0.4762         & AGN 2$^\ddagger$\\
N2\_52  & 16:36:29.71+41:02:22.7     & 16:36:29.78+41:02:23.0 & 1.44       & 0.03             & 37.8 $\times 10^{-15}$  & 1.009             & 0.02$^\dagger$ & AGN 1           \\
N2\_107 & 16:36:08.41+41:05:07.0     & 16:36:08.18+41:05:07.2 & 4.58       & 4.79             & 1.76 $\times 10^{-15}$  & 8.935             & 0.1683         & SB              \\
\hline
\end{tabular}
\caption{\label{tab:ids} {\it Chandra} sources with ISOCAM 15 $\mu m$
counterparts in the ELAIS regions N1 \& N2. $^1${\it Chandra} source
ID from Manners {\it et al.} (2003). $^2$Manners {\it et al.}
(2003). $^3$Rowan-Robinson {\it et al.} (2004). $^4$Offset between
{\it Chandra} and 15 $\mu m$ {\it ISO} positions. $^5$Probability of
this cross-correlation being a random association (see
section~\ref{sec:method}). $^6${\it Chandra} full band flux (Manners
{\it et al.} 2003). $^7$ISOCAM 15 $\mu m$ flux density (Rowan-Robinson
{\it et al.} 2004). $^8$Spectroscopic redshifts
($^\dagger$photometric, $1\sigma$ error $\sim 0.1$) reported in
Rowan-Robinson {\it et al.} (2003). $^9$Likely object class as
suggested by this paper (section~\ref{sec:ids}) or
$^\ddagger$previously confirmed by optical spectroscopy.  }
\end{table*}

\begin{table*}
\footnotesize
\begin{tabular}{l|ccccccc} \hline
                                    & N1\_20            & N1\_49                  &  N1\_69           & N2\_25            & N2\_33            &  N2\_52           & N2\_107\\
\hline
$f_{1.4 {\rm GHz}}$$^a$             & 0.26\spm0.01      & 5.08\spm0.01            & 0.15\spm0.01      & 0.13\spm0.01$^3$  & 57.62\spm0.02$^4$ &                   & 3.09\spm0.02$^4$\\
{\it ISO} $f_{175\mu{\rm m}}$$^a$   &                   &                         &                   &                   &                   &                   & 803\spm102\\
{\it IRAS} $f_{100\mu {\rm m}}$$^a$ &                   &                         &                   &                   &                   &                   & 898\spm126$^1$\\
{\it ISO} $f_{90\mu {\rm m}}$$^a$   &                   &                         &                   &                   &                   &                   & 614\spm37\\
{\it IRAS} $f_{60\mu {\rm m}}$$^a$  &                   &                         &                   &                   &                   &                   & 351\spm39$^1$\\
{\it ISO} $f_{15\mu {\rm m}}$$^a$   & 1.84\spm0.21      & 2.92\spm0.27            & 1.45\spm0.19      & 1.02\spm0.14      & 1.73\spm0.11      & 1.01\spm0.14      & 8.94\spm0.10\\
{\it ISO} $f_{6.7\mu {\rm m}}$ $^a$ &                   &                         &                   &                   &                   &                   & 2.30\spm0.12\\
K 2.2$\mu$m$^b$                     & 14.57\spm0.08     &                         &                   & 19.1\spm0.1$^3$   &                   & 13.04\spm0.07     & 14.05\spm0.12\\
H 1.65$\mu$m$^b$                    & 14.87\spm0.02$^2$ & 16.41\spm0.05$^\dagger$ & 15.91\spm0.03$^2$ & 21.20\spm0.03$^2$ & 16.23\spm0.03$^2$ & 13.33\spm0.05     & 14.61\spm0.14\\
J 1.25$\mu$m $^b$                   &                   &                         &                   &                   & 17.21\spm0.13     & 13.99\spm0.04     & 15.52\spm0.13\\
i$'$ 775nm$^b$                      & 16.64\spm0.02$^2$ &                         & 17.86\spm0.02$^2$ & 22.55\spm0.03$^2$ & 18.15\spm0.03$^2$ & 15.01\spm0.03$^2$ & 16.56\spm0.03$^2$\\
r$'$ 623nm$^b$                      & 17.12\spm0.02$^2$ &                         & 18.46\spm0.02$^2$ & 22.95\spm0.03$^2$ & 19.39\spm0.03$^2$ & 15.68\spm0.03$^2$ & 17.25\spm0.03$^2$\\
g$'$ 486nm$^b$                      & 17.68\spm0.02$^2$ & 23.88\spm0.13$^\dagger$ & 19.21\spm0.02$^2$ & 22.87\spm0.03$^2$ & 20.45\spm0.03$^2$ & 16.50\spm0.03$^2$ & 18.14\spm0.03$^2$\\
U 361nm $^b$                        & 17.61\spm0.02$^2$ & 23.78\spm0.16$^\dagger$ & 19.11\spm0.03$^2$ & 22.94\spm0.03$^2$ & 20.62\spm0.03$^2$ & 17.33\spm0.03$^2$ & 18.99\spm0.03$^2$\\
$f_{{\rm 0.5 - 2 keV}}$$^c$         & 0.8\spm0.3        & 1.0\spm0.3              & 23.0\spm1.2       & 2.1\spm0.4        & $<$ 0.5           & 14.2\spm1.0       & 1.0\spm0.3\\
$f_{{\rm 2 - 8 keV}}$$^c$           & $<$ 2.4           & 22.1\spm2.7             & 39.8\spm3.5       & 14.7\spm2.1       & 5.2\spm1.4        & 26.0\spm2.8       &  $<$ 3.9\\
\\
$\alpha_{IX}$                       & 1.53\spm0.03      & 1.32\spm0.01            & 1.20\spm0.01      & 1.26\spm0.02      & 1.41\spm0.03      & 1.21\spm0.02      & 1.70\spm0.03\\
HR                                  & -0.34\spm0.24     & 0.69\spm0.09            & -0.40\spm0.04     & 0.26\spm0.11      & 0.37\spm0.19      & -0.38\spm0.05     & -0.26\spm0.22\\
\hline
\end{tabular}
\caption{\label{tab:seds} Multi-waveband data for each
cross-correlated source. Uncertainties are $1\sigma$. X-ray data are
from Manners {\it et al.} (2003). Other waveband data are taken from
Rowan-Robinson {\it et al.} (2004) except $^1$Moshir {\it et al.}
(1990), $^2$Gonzalez-Solares {\it et al.} (in preparation),
$^3$Willott {\it et al.} (2003), $^4$Ciliegi {\it et al.} (1999),
$^\dagger$aperture magnitudes provided by E. Gonzalez-Solares. Also
included are the mid-IR to hard X-ray relation ($\alpha_{IX}$) and the
X-ray hardness ratio (HR) as defined in Manners {\it et al.}
(2003). Units are $^a$mJy, $^b$Vega magnitudes, and $^c$$10^{-15}$ erg
cm$^{-2}$ s$^{-1}$ }
\end{table*}

\begin{figure*}
\centering
\centerline{\epsfxsize=17 truecm \figinsert{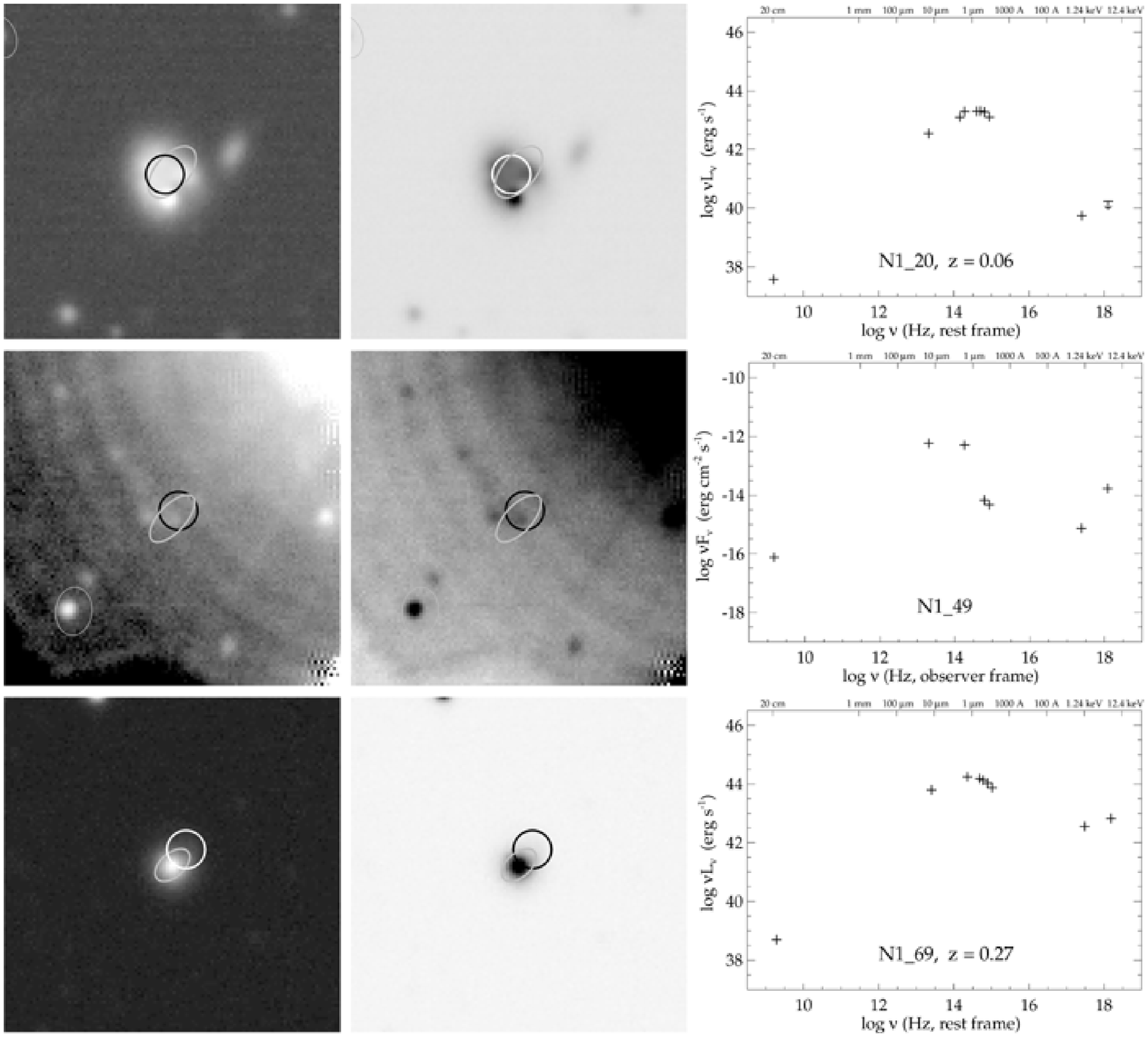}{0.0pt}}
\caption{{\it Chandra} sources with a 15 $\mu$m counterpart in the  N1
region.  25$''$ $\times$ 25$''$ r$'$-band postage stamps are from a
5400 second exposure with the INT WFC and are displayed in positive
and negative grey-scale  for clarity. Black or white circles indicate
position of the {\it ISO} source and  grey ellipses give the 3$\sigma$
size of the X-ray PSF.  Spectral energy distributions are at
rest-frame wavelengths and displayed in units of luminosity where
redshifts are available.
\label{fig:postage1}}
\end{figure*}

\begin{figure*}
\centering
\centerline{\epsfxsize=17 truecm \figinsert{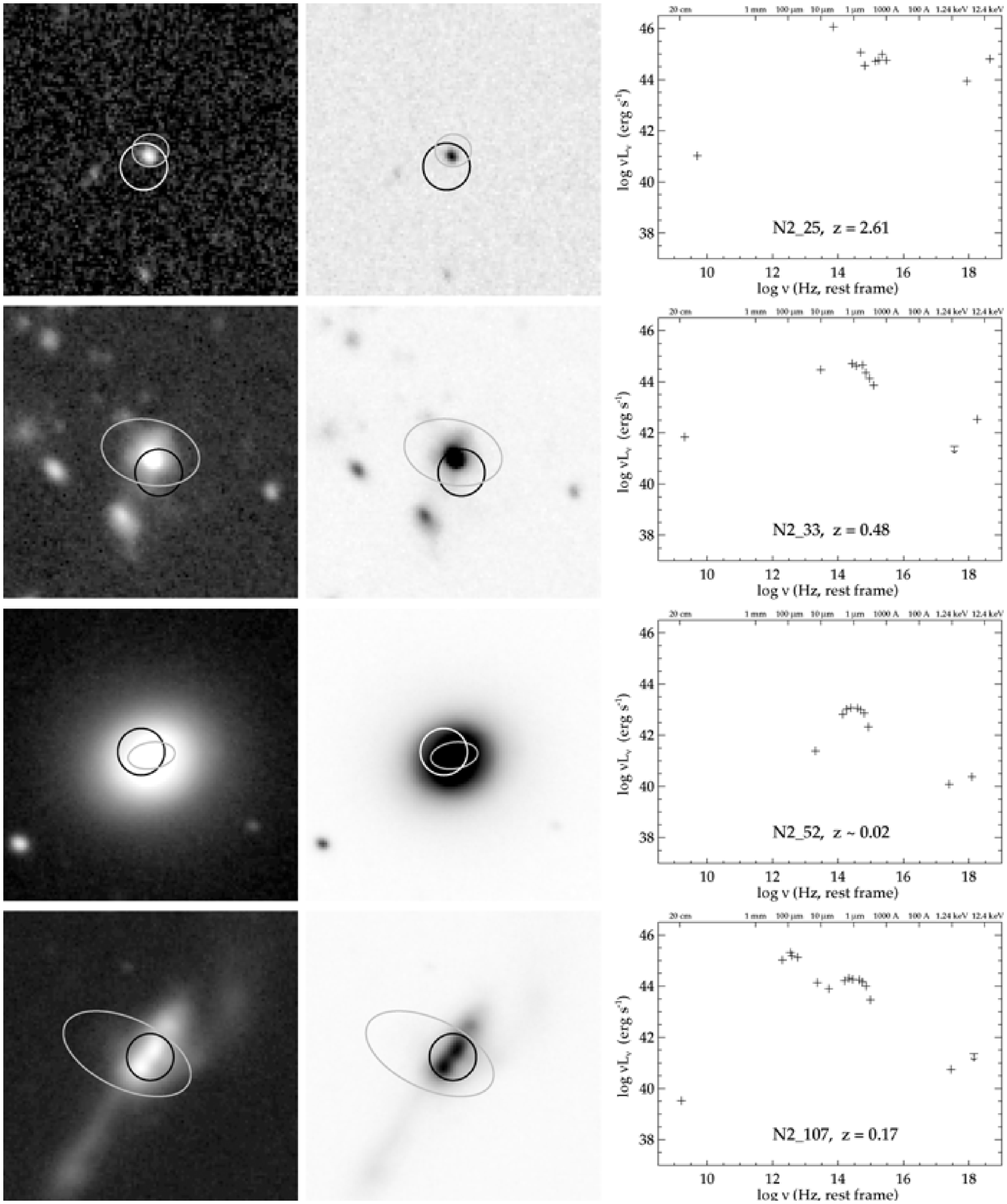}{0.0pt}}
\caption{{\it Chandra} sources with a 15 $\mu$m counterpart in the  N2
region.  25$''$ $\times$ 25$''$ r$'$-band postage stamps are from a
7800 second exposure with the WHT and are displayed in positive and
negative grey-scale  for clarity. Black or white circles indicate
position of the {\it ISO} source and  grey ellipses give the 3$\sigma$
size of the X-ray PSF.  Spectral energy distributions are at
rest-frame wavelengths and displayed in units of luminosity.
\label{fig:postage2}}
\end{figure*}

As part of ELAIS, ISOCAM observations were carried out covering 10.3
deg$^2$  at 15 $\mu$m. The observations were carried out in raster
mode, with most of  the survey area confined to three northern fields
(N1: 2.67 deg$^2$,  N2: 2.67 deg$^2$, N3: 0.88 deg$^2$) and one
southern field  (S1: 3.96 deg$^2$). Further details on the observation
strategy can be found  in Oliver {\it et al.} (2000).

Data reduction of ELAIS 15 $\mu$m observations was recently completed
by Vaccari {\it et al.} (in preparation). Data reduction was carried
out using the LARI method (Lari {\it et al.} 2001, Lari {\it et al.}
2003), a new technique devised for the reduction of ISOCAM and ISOPHOT
imaging data. Based on a physical model of the {\it ISO} detectors'
behaviour, the method is particularly suited for the reliable
detection of faint sources in {\it ISO} surveys, allowing sensitivity
to be pushed to the instrumental limits.

A sample of 1056 sources (490 in N1 and 566 in N2) were detected with
$S/N > 5$, spanning the 0.5 -- 100 mJy flux range and thus filling the
gap between the ISOCAM deep surveys (e.g. Elbaz {\it et al.} 1999) and
the {\it IRAS} Faint Source Catalogue (Moshir {\it et al.} 1990).

The ELAIS Deep X-ray Survey was centred on regions in the ELAIS N1
and N2 fields selected to have low cirrus and HI column density (see
Oliver {\it et al.} 2000 for more details).  The {\it Chandra}
observations are described in detail elsewhere (Manners {\it et  al.}
2003), though we summarise the main points here.  The observations
were taken with the {\it Chandra} ACIS (Advanced CCD Imaging
Spectrometer) array. Integrations of $75$ks were taken in each of N1
and N2. The pointing centroids are 16:10:20.11 +54:33:22.3 in N1 and
16:36:46.99 +41:01:33.7 in N2, with an area of $16.9' \times 16.9'$
covered by the ACIS-I chips in each case, giving  a total area of 571
sq. arcmin. The limiting flux levels are $4.6 \times 10^{-16}$ erg
s$^{-1}$ cm$^{-2}$ in the 0.5 - 2 keV band, and $2.2 \times 10^{-15}$
erg s$^{-1}$ cm$^{-2}$ in the 2 - 8 keV band.  The N1 region contains
125 {\it Chandra} sources from the 4 ACIS-I chips and 5 from the
ACIS-S2 chip. The N2 region has 99 sources on the ACIS-I chips and 4
on the ACIS-S2 chip. Only sources from the ACIS-I chips were used for
this analysis due to the poor resolution of the off-axis ACIS-S chips.

\section{Positional correlation between the {\it Chandra} and {\it ISO} sources}
\label{sec:method}
In the N1 region there are 9 $15\mu$m Final Analysis ELAIS sources
within the {\it Chandra} region, three of which are identified with
stars and the remainder of which are galaxy IDs.  In the N2 region
there are 24 $15\mu$m sources, two of which are identified with stars
and the rest with galaxies or blank fields in the r$'$ band image. The
positions of the 28 extragalactic sources are plotted in
Fig.~\ref{fig:sources} superimposed on the smoothed {\it Chandra}
images.

We performed a simple near-neighbour search to cross-correlate the
extragalactic {\it Chandra} and {\it ISO} sources within the area of
the  {\it Chandra} ACIS-I chips, using a $5''$ search
radius. Astrometric $1\sigma$ errors for the {\it Chandra} sources are
$\sim 1''$ (Manners {\it et  al.} 2003), while the nominal astrometric
accuracy for the ELAIS 15 $\mu$m FA catalogue sources ranges from
$\sim 0.8''$ to $2.0''$ (Vaccari {\it et al.} in preparation). The
search radius of $5''$ was chosen as the approximate sum of the $2\sigma$
astrometric errors. Three matches were found in the N1 region and four
in the N2 region (table \ref{tab:ids}, Fig.~\ref{fig:sources}). All
seven matches are with high reliability {\it ISO} sources (5$\sigma$).

To ensure the associations were real we calculated the probability
of a random association between each mid-IR source and its X-ray
counterpart. Following Fadda {\it et al.} (2002), we assume the X-ray 
counterpart belongs to a Poissonian distributed population so that
\begin{equation}
P = 1 - e^{-{\rm N(}>{\rm S)} \pi d^2}
\end{equation}
where P is the probability of a random association within an offset 
distance $d$. N($>$S) is the number density of sources with flux greater 
than the possible X-ray counterpart (S). We calculate this probability 
for each source (table \ref{tab:ids}) with reference to the log(N)-log(S) 
relation for these regions (Manners {\it et al.} 2003). The chances of random 
associations are found to be very low.

Figs.~\ref{fig:postage1} \&~\ref{fig:postage2} display the cross-correlations
overlaid on optical postage stamps.

\section{Results and discussion}\label{sec:results}

\subsection{Identifications}
\label{sec:ids}

Properties of the 7 cross-correlated sources are reported in
table~\ref{tab:ids}.  Multi-waveband data are reported and referenced in
table~\ref{tab:seds}. Also reported in table~\ref{tab:seds} are the 
mid-infrared to X-ray spectral indices ($\alpha_{IX}$, described in
section~\ref{sec:IRX}) and the X-ray hardness ratios (HR, from Manners
{\it et al.}  2003). r$'$ band postage stamps to a depth of r$'
\sim$26 are shown in Figs.  ~\ref{fig:postage1} \&~\ref{fig:postage2}
together with the spectral energy  distribution (SED) of each
source. For the 6 sources with available redshifts,  SEDs are
displayed in the rest frame in units of luminosity assuming a
cosmology with $\Omega_\Lambda = 0.73$, $\Omega_M = 0.27$, and H$_0 =
71$ km s$^{-1}$ Mpc$^{-1}$. For the source without a redshift
(N1\_49), the SED is displayed in the observed frame in units of
flux. In order to derive flux densities for the SEDs, magnitudes were
converted using the following zero points for each photometric band: K
657 Jy; H 1020 Jy; J 1600 Jy; i$'$ 2491 Jy; r$'$ 3133 Jy;
g$'$ 3876 Jy; U 1810 Jy. Flux densities at 1 keV and 5 keV were
derived from the 0.5 - 2 keV and 2 - 8 keV {\it Chandra} bands
respectively, assuming a power-law spectrum with photon index $\Gamma
= 1.7$ within each band. A description follows of the properties and
most probable identifications for  each source:

\begin{description}
\item [{\bf N1\_20 (CXOEN1 J161046.5+543538)}] A complex r$'$ band
morphology indicative of a recent merger. This low redshift
($z=0.0634$) object has a soft X-ray  spectrum and is below the
detection threshold in the 2 - 8 keV band image. The  mid-infrared to
X-ray spectral index is also quite steep ($\alpha_{IX}=1.51$).  These
characteristics are consistent with identification as a starburst
galaxy (SB, table~\ref{tab:ids}).
\item [{\bf N1\_49 (CXOEN1 J161020.8+543900)}]  Spectroscopic
identification of this object is hampered by the presence of a  nearby
bright star (as can be seen in the r$'$ band postage stamp). No
redshift  is available, however a very hard X-ray spectrum combined
with a relatively flat  mid-infrared to X-ray spectral index indicates
likely identification with an AGN.  The X-ray spectrum is consistent
with an apparent absorbing column of N$_H \sim 6 \pm 2 \times 10^{22}$
cm$^{-2}$ (assuming an underlying power  law of $\alpha = 0.7$ and $z
= 0$). Depending on the redshift of this source  the actual N$_H$ is
likely to be higher with apparent absorbing column  scaling as
$(1+z)^{2.6}$. The mid-infrared to X-ray spectral index
($\alpha_{IX}=1.32$) is also consistent with an absorbed AGN (see
section ~\ref{sec:IRX}). Given an AGN 2 classification in
table~\ref{tab:ids}.
\item [{\bf N1\_69 (CXOEN1 J161003.1+543628)}] Spectroscopic redshift
of 0.2675. The flat mid-infrared to X-ray spectral index,  luminosity
and X-ray hardness ratio are all typical of a Seyfert galaxy (AGN 1,
table~\ref{tab:ids}).
\item [{\bf N2\_25 (CXOEN2 J163655.7+405910)}] Spectroscopically
confirmed quasar at a redshift of 2.61 (AGN 1, table~\ref{tab:ids}). 
Its properties are  extensively reported in Willott {\it et al.} (2003).
\item [{\bf N2\_33 (CXOEN2 J163651.6+405600)}]    Spectroscopically
identified as a Seyfert 2 galaxy at a redshift of 0.4762 (AGN 2,
table~\ref{tab:ids}). The  r$'$ band image indicates an interaction is
occurring with a smaller neighbour.  Radio morphology suggests this to
be an FR-II type AGN.
\item [{\bf N2\_52 (CXOEN2 J163629.7+410222)}] Elliptical galaxy with
no obvious point-like nuclear source. A photometric redshift of $\sim
0.02$ suggests an X-ray luminosity that is low for an active galaxy
($1\times 10^{40}$ erg s$^{-1}$ at 1 keV, $2\times 10^{40}$ erg
s$^{-1}$ at 5 keV). However, mid-infrared to X-ray spectral index and
X-ray hardness ratio are both consistent with an AGN (AGN 1,
table~\ref{tab:ids}).
\item [{\bf N2\_107 (CXOEN2 J163608.4+410507)}] Interesting r$'$ band
morphology displaying merging galaxies with a double
nucleus. Spectroscopic redshift of 0.1683. The X-ray spectrum is soft
and of low-luminosity ($6\times 10^{40}$ erg s$^{-1}$ at 1 keV) with
the source  undetected in the 2 - 8 keV band image. The relatively
luminous infra-red  spectrum supports identification with a starburst
galaxy (SB, table~\ref{tab:ids}).
\end{description}

Of the 7 matched sources, 1 is a spectroscopically confirmed quasar, 1
is spectroscopically identified as a Seyfert 2, a further 3 display
properties of AGN, and 2 have properties consistent with starburst galaxies.

\subsection{The mid-infrared to X-ray spectral indices}
\label{sec:IRX}

\begin{figure}
\centering
\centerline{\epsfxsize=8.5 truecm \figinsert{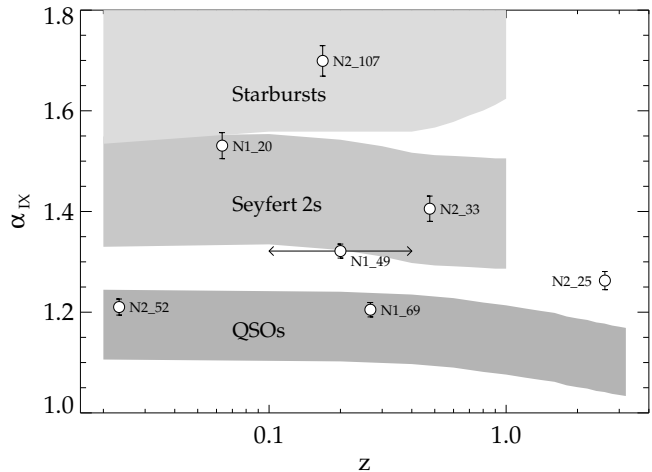}{0.0pt}}
\caption{Mid-infrared to hard X-ray spectral indices ($\alpha_{IX}$) vs. 
redshift. Shaded areas represent the error boxes for template SEDs of QSOs, 
Seyfert 2s and Starburst galaxies as reported in Alexander {\it et al.} 
(2001, Fig. 7). cf. Fadda {\it et al.} (2002, Fig. 8).
\label{fig:xir}}
\end{figure}

Values for the mid-infrared to X-ray spectral index ($\alpha_{IX}$) are 
calculated using the flux density observed at 15$\mu$m and 5 keV,
assuming a power-law spectrum of the form F$_\nu \propto
\nu^{-\alpha_{IX}}$. The flux density at 5 keV is calculated from flux
in the 2 - 8 keV {\it Chandra} band. For the 2 objects undetected in this
band, the full 0.5 - 8 keV band is used.

The mid-infrared to X-ray spectral index can be a useful indicator to
distinguish between AGN and starburst galaxies (e.g. Alexander {\it et
al.}  2001, Fadda {\it et al.} 2002). Starburst galaxies are found to
have high values of  $\alpha_{IX}$, while type-1 AGN have low values
($\alpha_{IX} < 1.2$).  Type-2 AGN have values in between depending on
the amount of obscuration.  Fig.~\ref{fig:xir} plots the $\alpha_{IX}$
values for our 6 matched sources with available redshifts. The figure
displays values of $\alpha_{IX}$ as a function of redshift for
template  SEDs compiled by Alexander {\it et al.} (2001). Source
N2\_33, spectroscopically identified as a Seyfert 2 lies squarely
within the region of the Seyfert 2 templates. N2\_25,
spectroscopically identified as a QSO has a value slightly higher than
the QSO templates although still lower than expected for a Seyfert
2. Of the remaining objects studied here, N1\_69 and N2\_52 lie in the
region of type-1 AGN, N1\_20 and N2\_107 are consistent with
starbursts or highly  obscured AGN, and N1\_49 is  consistent with a
type-2 AGN.

\subsection{Constraints on models of mid-infrared source counts} 

In the N1 region there are 6 extragalactic $15\mu$m Final Analysis
ELAIS sources within the {\it Chandra} region.  In the N2 region
conversely there are 22 extragalactic $15\mu$m sources. The {\it
Chandra} N2 region falls within an area of repeated {\it ISO}
observations while the {\it Chandra} N1 region does not. The
difference in source counts, however, is larger than expected and may
indicate clustering on scales larger than the field size. This
highlights the need for larger areas to be surveyed before sufficient
count statistics can be gained. Here we consider our results along
with number counts from the Hubble Deep Field - North (HDF-N) and the
Lockman Hole (Fadda {\it et al.} 2002) in order to provide a
comparison with recent models of mid-infrared source counts.

The $15\mu$m flux range $0.8 - 6.0$ mJy is well covered by the {\it
ISO} observations in our {\it Chandra} regions. In this flux range
there are 26 extragalactic $15\mu$m sources in our sample.  We take 5
of these objects to contain evidence of AGN from the {\it Chandra} and
multi-waveband data. This gives an AGN fraction of 5/26 extragalactic
sources, or $0.19\pm0.09$ over the given flux range, in a total area
of $\sim 571$ sq. arcmin. This fraction will be a lower limit if any
heavily obscured, Compton-thick AGN are present in our sample.

Fadda {\it et al.} (2002) report AGN number counts for mid-infrared
sources in the Lockman Hole and HDF-N. They find for the Lockman Hole
an AGN fraction of 13/103 extragalactic sources ($0.13\pm0.04$) within the
$15\mu$m flux range $0.5 - 3.0$ mJy in a total area of 218
sq. arcmin. For the HDF-N the AGN fraction is 5/42 ($0.12\pm0.05$) over
a $15\mu$m flux range of $0.1 - 0.5$ mJy in a total area of 24.3
sq. arcmin.

\begin{figure}
\centering
\centerline{\epsfxsize=8.5 truecm \figinsert{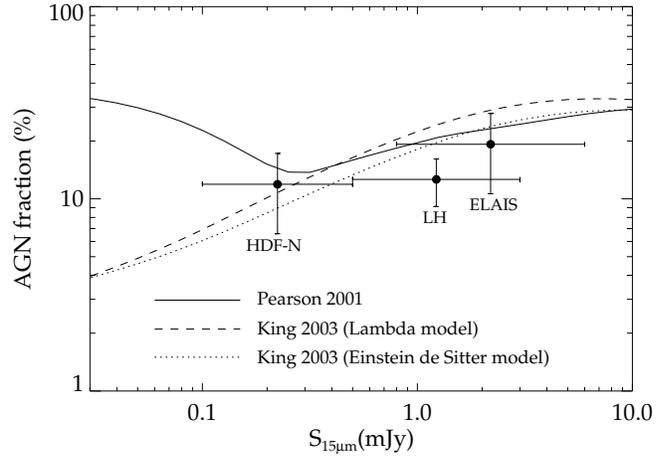}{0.0pt}}
\caption{ AGN fraction in the mid-infrared population derived from the
ELAIS Deep X-ray survey (this work), as well as fractions derived by
Fadda {\it et al.} (2002) from {\it Chandra} observations of the
Hubble Deep Field North (HDF-N) and {\it XMM-Newton} observations of the
Lockman Hole (LH). Also plotted are  the predictions from the integral
counts of Pearson (2001), and two models  from King \& Rowan-Robinson
(2003) (based on the models of Rowan-Robinson 2001).
\label{fig:agn_fraction}
}
\end{figure}

We compare these results with models for the AGN contribution to
mid-infrared surveys by Pearson (2001) and King \& Rowan-Robinson
(2003).  These models primarily attempt to model the star formation
history by fitting infrared source count observations with
contributions from normal galaxies, starbursts, ultra-luminous infrared
galaxies (ULIRGs) and AGN. The Pearson (2001) model uses local
luminosity functions and pure luminosity evolution to describe the
normal galaxy, starburst and AGN populations while the ULIRG component
is evolved in both density and luminosity. The $15\mu$m and $850\mu$m
source counts are used to constrain the model. The best fit to the
observations is obtained where the ULIRG component undergoes two major
phases of evolution, rapid merging to $z \sim 1$ and an exponential
evolution in luminosity to higher redshifts. This has an effect on the
AGN fraction of the source counts which will be tested here. The King
\& Rowan-Robinson (2003) models are updated versions of models
developed by Rowan-Robinson (2001) and are calculated for lambda and
Einstein de Sitter cosmologies. They allow for both density and
luminosity evolution in all four populations. The AGN contribution in
both models is found using a $12\mu$m luminosity function from Rush,
Malkan \& Spinoglio (1993) which does not include Compton-thick AGN
and can therefore be directly compared with these observations.
Fig.~\ref{fig:agn_fraction} displays the AGN fraction predicted by
these models against source flux. All 3 models shown are consistent
with AGN fractions reported here for the ELAIS regions and for the
HDF-N. The AGN fraction in the Lockman Hole is slightly lower than the
predictions of these models.

\begin{figure}
\centering
\centerline{\epsfxsize=8.5 truecm \figinsert{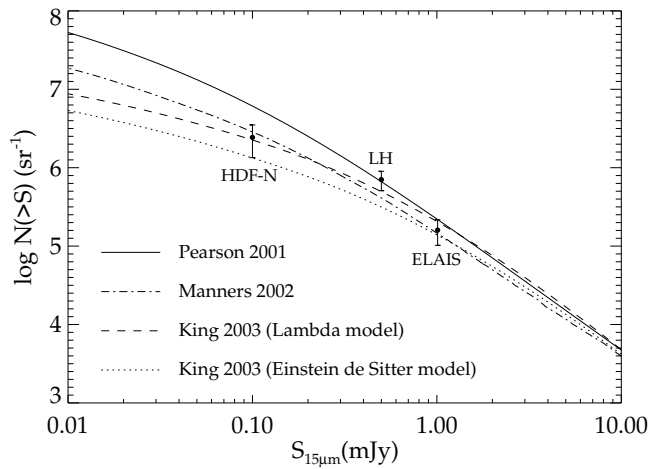}{0.0pt}}
\caption{Integral AGN number counts for $15\mu$m surveys. Values for 
the Lockman Hole and HDF-N are taken from Fadda {\it et al.} (2002). Number 
count predictions are taken from the same models displayed in 
Fig.~\ref{fig:agn_fraction}, plus that of Manners (2002).
\label{fig:iragn}
}
\end{figure}

We also compare the observations to modelled predictions for the
cumulative AGN source counts at $15\mu$m (Fig.~\ref{fig:iragn}).  To
make a reliable comparison it was necessary to correct our AGN source
counts for completeness. While the 15 $\mu$m data reduction has been
completed by Vaccari {\it et al.} (in preparation), completeness
estimates are to be published in Lari {\it et al.} (in preparation)
and are not yet available. Therefore, values from Gruppioni {\it et
al.} (2002) were taken as representative. By estimating the
completeness at the flux level of our sources we calculate the
expected number of AGN above the flux limit to be $\sim 7.7$ (ELAIS
data point, Fig.~\ref{fig:iragn}).  We can now compare a further model
from Manners (2002) which assumes AGN are responsible for the entire
hard X-ray background and predicts the emission of these AGN in the
mid-infrared. Observations in the HDF-N agree with models by King \&
Rowan-Robinson (2003) and Manners (2002), while the Pearson (2001)
model over-predicts the number of sources at this depth. AGN counts
from the Lockman Hole are consistent with the Pearson (2001) and King
\& Rowan-Robinson (2003, Lambda) models, whilst the AGN counts
reported here for the ELAIS survey agree well with all the model
predictions.

\subsection{Statistical hard X-ray limits of ELAIS {\it ISO} sources}

One further method of constraining the source count models is the
statistical detection of {\it ISO} sources in the {\it Chandra}
map. By co-adding the {\it Chandra} fluxes at the positions of {\it
ISO} sources not detected individually, we can obtain a constraint on
the mean hard X-ray flux and hard X-ray to mid-infrared flux ratio for
the remaining population.

In the N1 region there are 3 extragalactic sources undetected by {\it
Chandra}  whilst in the N2 region there are 18 (see
Fig.~\ref{fig:sources}). In order  to obtain limits on the X-ray flux
for these sources it was necessary to define  source regions on the
{\it Chandra} images. These regions were positioned accurately  using
the astrometry obtained by cross-correlating {\it Chandra} and
r$'$-band sources (Manners {\it et al.} 2003). The size of each region
was defined as a circle containing 95\% encircled energy for a
monoenergetic {\it Chandra} PSF at 4.51 keV at  the relevant source
position.

For each field, around 6 circular background regions of radii ranging
from $\sim 1' - 2'$ were selected. These were chosen carefully to
avoid contamination from known {\it Chandra} sources and
to cover regions of typical effective exposure. The background counts
expected in each source region were then calculated by correcting for
the difference in effective exposure. This method was preferred over
the use of individual local background estimates due to the slow
change in background counts over the image and the better statistics
gained from using the larger background regions.

Counts were extracted from the hard (2 - 8 keV) and soft (0.5 - 2 keV)
band  {\it Chandra} images. Source regions were individually corrected
for effective exposure,  background subtracted, and then co-added to
obtain an estimate of the total flux.

In the hard band we obtain a total of 58 counts with 49.3 background
counts expected. The Poisson probability of obtaining 58 or more
counts with 49.3 expected is 0.123, equivalent to a tentative
detection at a confidence level of 87.7\%.  This is equivalent to  a
mean flux of 2.2 $\pm$ 1.5 $\times 10^{-16}$ erg cm$^{-2}$ s$^{-1}$
for each of the 21 sources.  In the soft band we obtain a total of 43
counts with 24.1 background counts expected. The Poisson probability
of obtaining 43 or more counts with 24.1 expected is 3 $\times
10^{-4}$, equivalent to a reliable detection at a confidence level of
99.97\%. These counts are equivalent to a mean flux of 6.7 $\pm$ 2.5 
$\times 10^{-17}$ erg cm$^{-2}$ s$^{-1}$.

In order to check the validity of the stacking analysis we repeated
the procedure with the positions randomised over the areas of the {\it
Chandra} field that are free of known sources. This was repeated 10
times for each band. The distributions of the counts obtained were
consistent with a Gaussian of mean and variance equal to the expected
number of background counts in each case.

The stronger detection in the soft band is indicative of the higher
efficiency of {\it Chandra} in this band and does not imply these
sources have a particularly soft X-ray spectrum. The hardness ratio
(defined in Manners {\it et al.} 2003) for  the co-added regions is
-0.09 $\pm$ 0.38, somewhat harder than expected for
starbursts. However, the low significance of the hard X-ray detection
means this must be used with caution.

Seven of the unmatched sources have spectroscopic redshifts available,
ranging from $z = 0.10$ to 0.24. At a typical redshift of 0.2 the mean
fluxes given above would equate to luminosities of $5.4 \times
10^{39}$ at 1 keV and $1.9 \times 10^{40}$ at 5 keV. This is entirely
consistent with a starburst origin for the X-ray emission.  The mean
$15\mu$m flux for the 21 unmatched sources is 1.489 mJy. This leads to
a  mean spectral index of $\alpha_{IX} = 1.68$, also consistent with
purely starburst galaxies.

It should be noted that `Compton thick' AGN will not appear in the
X-ray data and constitute an unknown fraction of the unmatched sources.

\subsection{The cosmic star formation history from mid-infrared flux limited samples}

Of the sources detected in field surveys by the {\it ISO} satellite, by
far the largest fraction and largest number of moderate-redshift
($z>0.5$) systems are found in mid-infrared ISOCAM surveys, as opposed
to the far-infrared ($90-175\mu$m) surveys conducted by the ISOPHOT
instrument. Also, the surveys now being performed by the {\it Spitzer}
satellite, such as SWIRE (Lonsdale {\it et al.} 2003), will have
high-redshift objects preferentially detected in the mid-infrared
passbands. The AGN torus emission peaks in the infrared (e.g. Haas
{\it et al.} 1998) so it
might be expected that a large population of AGN would
occur in these surveys in addition to high-$z$ star forming galaxies.
Our AGN fractions, together with other studies (Fadda {\it et al.}
2002, Alexander {\it et al.} 2002), show this population to be a fairly
well-determined minority and are encouraging for the use of
mid-infrared samples for constraining the 
cosmic star formation history. The mid-infrared luminosity is a
reasonably good star formation rate indicator, albeit affected by
complicated K-correction effects (e.g. Xu {\it et al.} 1998).

The main caveat is that our {\it Chandra} data still do not
exclude the possibility of a large population of Compton-thick objects
at moderate  redshifts ($z\sim1$). Alexander {\it et al.} (2002),
however, argue that this fraction should be low. They performed an
X-ray stacking analysis on those infrared galaxies that were not
clearly AGN at X-ray energies but were still individually detected in
the 1 Ms Chandra Deep Field North Survey. They find an average X-ray
spectral slope of $\Gamma = 2.0$, indicating a low contribution from
the much flatter spectra of obscured AGN.

\section{Conclusions}

We performed a cross-correlation of X-ray and mid-infrared point
sources in the ELAIS N1 and N2 fields, using data from the  {\it
Chandra} and {\it ISO} satellites. 7 extragalactic matches are found
(out of a total of 28 {\it ISO} sources) within the area of the {\it
Chandra} ACIS-I chips.  2 of these are spectroscopically identified as
AGN. Based on X-ray to IR flux ratios, X-ray hardness ratios, and
luminosities, 3 of the remaining 5 are also consistent with AGN while
the other 2 are consistent with starburst galaxies. This provides an
AGN fraction of $\sim 19$ per cent in the $15\mu$m flux range $0.8 -
6$ mJy. We have co-added the hard X-ray flux at the positions of the
21 undetected {\it ISO} sources providing only a 1.4 sigma
detection. This translates to a mean hard X-ray to mid-IR flux ratio
consistent with star formation in these objects.  Our
cross-correlations, when compared with {\it XMM-Newton} observations
of the Lockman Hole and {\it Chandra} observations of the northern
Hubble Deep Field, allow us to place constraints on source count
models of the $15\mu$m source population.  The AGN fractions and
number counts are broadly consistent with models by Pearson (2001) and
King \& Rowan-Robinson (2003).  We argue our data is encouraging for
the use of mid-infrared samples to constrain the cosmic star formation
history, provided there is not a large contribution from Compton-thick AGN.

\section*{Acknowledgements}

This research has made use of the NASA Extra-galactic Database (NED),
operated by the Jet Propulsion Laboratory, California Institute of
Technology, under contract with the National Aeronautics and Space
Administration. This work was partly funded under PPARC grant number
GR/K98728. JCM would like to thank the referee for detailed and useful
comments.

\end{document}